\documentclass[aps,prl,twocolumn,longbibliography] {revtex4-1}
\usepackage{amsmath,graphicx,amsfonts}

\usepackage{times}
\usepackage[colorlinks=true,linkcolor=blue,urlcolor=blue,citecolor=blue, breaklinks=true,hypertexnames=false]{hyperref}

\usepackage[utf8]{inputenc}
\DeclareUnicodeCharacter{03BC}{\mbox{$\mu$}}

\DeclareUnicodeCharacter{03B4}{$\delta$}
\DeclareUnicodeCharacter{2009}{-}

\DeclareMathOperator{\sech}{sech}

\usepackage[english]{babel}
\addto\captionsenglish{}

\makeatletter
\def\bbl@set@language#1{%
	\edef\languagename{%
		\ifnum\escapechar=\expandafter`\string#1\@empty
		\else\string#1\@empty\fi}%
	\@ifundefined{babel@language@alias@\languagename}{}{%
		\edef\languagename{\@nameuse{babel@language@alias@\languagename}}%
	}%
	\select@language{\languagename}%
	\expandafter\ifx\csname date\languagename\endcsname\relax\else
	\if@filesw
	\protected@write\@auxout{}{\string\select@language{\languagename}}%
	\bbl@for\bbl@tempa\BabelContentsFiles{%
		\addtocontents{\bbl@tempa}{\xstring\select@language{\languagename}}}%
	\bbl@usehooks{write}{}%
	\fi
	\fi}
\newcommand{\DeclareLanguageAlias}[2]{%
	\global\@namedef{babel@language@alias@#1}{#2}%
}
\makeatother

\DeclareLanguageAlias{en}{english}

\makeatletter
\def\@bibdataout@aps{%
	\immediate\write\@bibdataout{%
		@CONTROL{%
			apsrev41Control%
			\longbibliography@sw{%
				,author="08",editor="1",pages="1",title="0",year="1"%
			}{%
				,author="08",editor="1",pages="1",title="",year="1"%
			}%
		}%
	}%
	\if@filesw \immediate \write \@auxout {\string \citation {apsrev41Control}}\fi 
}
\makeatother

\def\TITLE{Quasiparticle energy relaxation in a gas of one-dimensional fermions with Coulomb interaction}

\date\today
\begin{document}
	
\title{\TITLE}
\author{Zoran Ristivojevic$^1$ and K. A. Matveev$^2$}
\affiliation{$^1$Laboratoire de Physique Th\'{e}orique, Universit\'{e} de Toulouse, CNRS, UPS, 31062 Toulouse, France}
\affiliation{$^2$Materials Science Division, Argonne National Laboratory, Argonne, Illinois 60439, USA}
	
\begin{abstract}
We consider a  system of charged one-dimensional spin-$\frac{1}{2}$ fermions at low temperature. We study how the energy of a highly-excited quasiparticle (or hole) relaxes toward the chemical potential in the regime of weak interactions. The dominant relaxation processes involve collisions with two other fermions. We find a dramatic enhancement of the relaxation rate at low energies, with the rate scaling as the inverse sixth power of the excitation energy. This behavior is caused by the long-range nature of the Coulomb interaction.
\end{abstract}

\maketitle

The Tomonaga-Luttinger liquid theory is widely used to describe low-energy properties of interacting fermions in one dimension \cite{Giamarchi}. It is based on the model of interacting  fermions with linear dispersion, which admits an exact solution. The resulting excitation spectrum is that of a system of noninteracting bosons \cite{mattis_exact_1965}. This idealization is appropriate in the low-energy limit. Importantly, this model is free of inelastic scattering and thus it cannot describe relaxation of the system towards equilibrium.

Recent theoretical progress has shown the importance of the nonlinear corrections to the spectrum, as they affect response functions and enable quasiparticle relaxation \cite{imambekov_one-dimensional_2012,levchenko_kinetic_2021}. Experiments with one-dimensional conductors support these findings. In particular, the behavior of the response functions was probed in Refs.~\cite{jin_momentum-dependent_2019,wang_nonlinear_2020}, equilibration rates for hot electrons and holes were measured in Ref.~\cite{barak_interacting_2010}, while peculiar features of the relaxation of very hot electrons were observed in Ref.~\cite{reiner_hot_2017}. These experiments have demonstrated the crucial role of the curvature of the spectrum of electrons.

\begin{figure}[b]
	\includegraphics[width=\columnwidth]{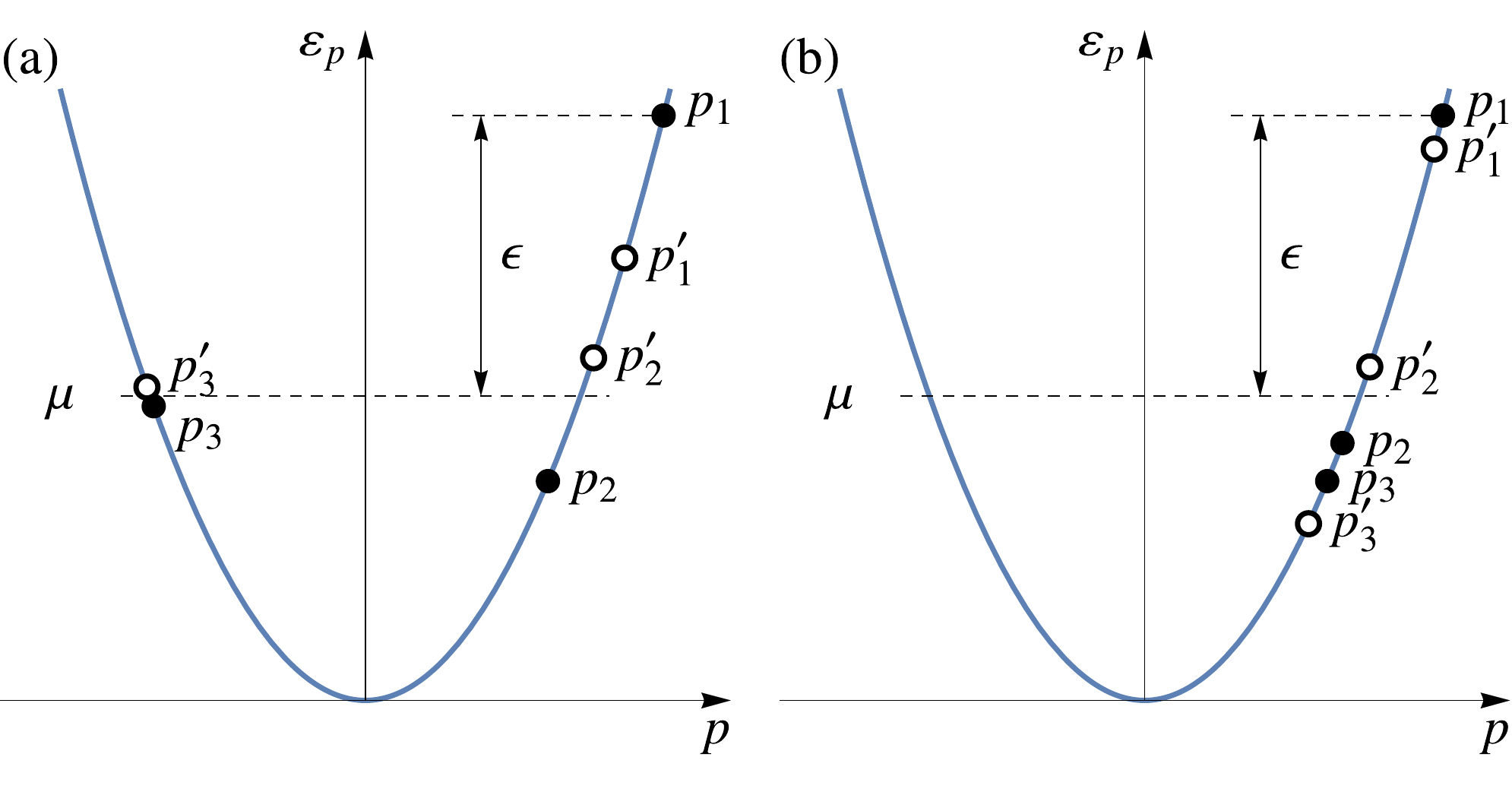}
	\caption{Different scattering mechanisms that contribute to  relaxation  of quasiparticles in a one-dimensional system of weakly-interacting fermions.  At $T=0$, only the processes of type (a) are allowed. At nonzero temperature, the processes of type (b) are responsible for the dominant contribution to the relaxation rate at energies $\epsilon \ll \sqrt{T\mu}$.} \label{fig:hole}
\end{figure}

Significant theoretical progress has been achieved in the case of weakly interacting fermions with quadratic spectrum \cite{khodas_fermi-luttinger_2007,imambekov_one-dimensional_2012}. In one dimension, pair collisions result in identical sets of momenta before and after scattering. As a result, the decay of quasiparticles is controlled by three-particle scattering processes \cite{lunde_three-particle_2007}. For quasiparticles with energies near the Fermi level, the two types of processes shown in Fig.~\ref{fig:hole} should be considered. In  the  initial state, the scattering processes of type (a) have one particle with the opposite sign of momentum than the other two, while all three particles are near the same Fermi point for the processes of type (b). Due to the conservation laws, the final states of the three particles are in the same configuration as the initial ones. It is worth noting that the processes of type (b) are allowed only at finite temperature $T$, whereas those of type (a) bring about the relaxation of quasiparticles even at $T=0$	 \cite{khodas_fermi-luttinger_2007}.

Relaxation of quasiparticles in the system of spin-$\frac{1}{2}$ fermions with weak Coulomb repulsion was considered in Ref.~\cite{karzig_energy_2010}. At zero temperature, a quasiparticle with the energy $\epsilon$ above the Fermi level decays with the rate $\tau^{-1}\propto \epsilon^2$ \footnote{Here we have neglected the factors that scale logarithmically with energy.}. At finite temperatures this result applies as long as $\epsilon\gg\sqrt{T\mu}$, where $\mu$ is the chemical potential of the Fermi gas. At energies below $\sqrt{T\mu}$ the  quasiparticle relaxation rate was found to have only a weak dependence on energy, $\tau^{-1} \propto \ln^2(\mu/\epsilon) T$. Both rates are due to the processes shown in Fig.~\ref{fig:hole}(a) \footnote{Note that the decay rates for spinless fermions are very different, as they scale with higher powers of $\epsilon$ or $T$	\cite{levchenko_kinetic_2021,khodas_fermi-luttinger_2007,micklitz_thermalization_2011,ristivojevic_relaxation_2013,matveev_decay_2013,protopopov_relaxation_2014,protopopov_equilibration_2015}.}. \nocite{khodas_fermi-luttinger_2007,micklitz_thermalization_2011,ristivojevic_relaxation_2013,matveev_decay_2013,protopopov_relaxation_2014,protopopov_equilibration_2015}

\begin{figure}
	\includegraphics[width=0.8\columnwidth]{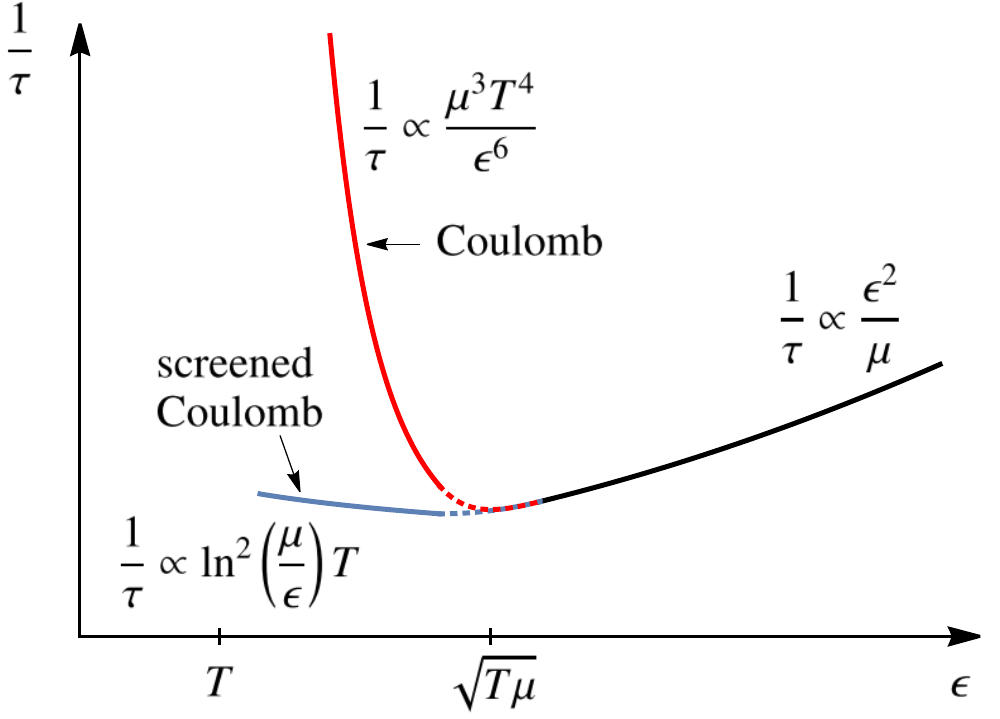}
	\caption{Sketch of the energy dependence of the quasiparticle relaxation rate for spin-$\frac{1}{2}$ fermions with Coulomb and screened Coulomb interactions. In the former case there is a rapid increase of the rate at energies below $\sqrt{T \mu}$ as opposed to a gradual logarithmic rise in the latter case. A similar sharp increase of the relaxation rate at low energies also occurs for holes.} \label{fig3}
\end{figure}

It is important to note that in Ref.~\cite{karzig_energy_2010} the Coulomb interaction was assumed to be screened at small momentum transfers by a nearby gate, which enabled the authors to neglect the contribution of type (b) processes to the relaxation rate. In this paper we show that type (b) processes lead to a dramatically different behavior in the unscreened case. We found that at quasiparticle energies below $\sqrt{T\mu}$ it gives the dominant contribution to the relaxation rate, which behaves as $\tau_b^{-1}\propto \mu^3 T^4/\epsilon^6$, see Fig.~\ref{fig3} \cite{Note1}. This implies a drastic enhancement of the  rate as the quasiparticle excitation energy $\epsilon$ drops below the characteristic energy $\sqrt{T\mu}$, in contrast to weak energy dependence $\tau^{-1} \propto \ln^2(\mu/\epsilon) T$ for the screened case \cite{karzig_energy_2010}. This behavior is qualitatively different from that of quasiparticles in most other systems of fermions, where the relaxation rate decreases at lower energies.  For example, in three-dimensional Fermi liquids $\tau^{-1}\propto\epsilon^2$ \cite{pines}.

We study a one-dimensional system of fermions with quadratic dispersion $\varepsilon_p=p^2/2m$ and weak two-body interaction. In second quantization, the latter is described by
\begin{align}\label{eq:V}
\hat{V}=\frac{1}{2L}\sum_{{p_1,p_2,q\atop \sigma_1,\sigma_2}}V(q)  \hat a_{p_1+q,\sigma_1}^\dagger \hat a_{p_2-q,\sigma_2}^\dagger \hat a_{p_2,\sigma_2} \hat a_{p_1,\sigma_1}. 
\end{align}
Here $\hat a$ and $\hat a^\dagger$ are the fermionic spin-$\frac{1}{2}$ operators obeying the standard anti-commutation relations, $L$ is the system size, while $V(q)$ is the Fourier transform of the two body interaction potential. For electrons in a quantum wire the latter has the Coulomb form $U(x)=e^2/|x|$ that should be cut off at short distances by the width of the wire $w$. Here $e$ denotes the electron charge. At small momenta $|q|\ll \hbar/w$, the Fourier transform of the interaction potential is $V(q)=2e^2 \ln\left(\hbar/|q|w\right)$. 

Let us consider a right-moving quasiparticle well above the Fermi level, i.e., with energy $\epsilon=\varepsilon_p-\mu\gg T$, where $p$ denotes the quasiparticle momentum. Such an energetic quasiparticle on average loses its energy due to collisions with other quasiparticles and thus drifts towards the Fermi level. The relaxation proceeds predominantly via three-particle scattering processes where the other two quasiparticles are near the Fermi level. In this case the rate of energy change of the initial quasiparticle is given by \begin{align}\label{eq:tau}
\dot{\epsilon}={}&\frac{1}{2}\sum_{{p_1>p_2>p_3\atop p_1'>p_2'>p_3'}}\left(\varepsilon_{p_1'}-\varepsilon_{p_1}\right) W_{p_1,p_2,p_3}^{p_1',p_2',p_3'}\notag\\
&\times n_{p_2}n_{p_3}(1-n_{p_1'})(1- n_{p_2'})(1- n_{p_3'})\delta_{p,p_1}.
\end{align}
Here $W_{p_1,p_2,p_3}^{p_1',p_2',p_3'}$ is the scattering rate of the three fermions with momenta $p_1,p_2,p_3$ into $p_1',p_2',p_3'$ summed over all spin indices, while $n_p$ denotes the Fermi distribution function. The prefactor $1/2$ in Eq.~(\ref{eq:tau}) compensates for the summation over the spin of the initial quasiparticle. The main focus of this paper is the quasiparticle relaxation that arises due to processes shown in Fig.~\ref{fig:hole}(b). In this case all the momenta that participate in the sum of Eq.~(\ref{eq:tau}) are positive.

The conservation laws of momentum and energy enable us to estimate the momentum change of the initial quasiparticle in a three-particle collision. For quadratic dispersion we find 
\begin{align}\label{eq:exact}
p_1-p_1'=\frac{(p_3'-p_3)(p_3'-p_2)}{p_1-p_2'}.
\end{align} 
For the typical processes shown in Fig.~\ref{fig:hole}(b), the momenta $p_2$, $p_3$, and $p_3'$ are near the Fermi point, and $|p_3'-p_3|,|p_3'-p_2|\sim T/v_F$, where $v_F$ is the Fermi velocity. In combination with the momentum conservation law, this yields
\begin{align}\label{eq:deltaeps1}
v_F|p_1-p_1'|\sim \frac{T^2}{\epsilon}\ll \epsilon.
\end{align}
Thus, for type (b) processes, both the initial and final states have one highly excited quasiparticle, while the other two are always near the Fermi level. This enables us to identify the fermion at $p_1'$ as a new state of the initial quasiparticle after the scattering event. Equation (\ref{eq:tau}) shows how the energy of this quasiparticle changes with time. We define
\begin{align}\label{eq:taudef}
\frac{1}{\tau}=-\frac{\dot{\epsilon}}{\epsilon}
\end{align}
as the energy relaxation rate. In this paper we distinguish it from the quasiparticle decay rate, which is obtained by omitting $(\varepsilon_{p_1'}-\varepsilon_{p_1})$ in Eq.~(\ref{eq:tau}).

For the processes shown in Fig.~\ref{fig:hole}(a), after the scattering event the two right-moving quasiparticles have energies on the order of $\epsilon$ \cite{karzig_energy_2010}. This is qualitatively different from the case of type (b) processes, where only one quasiparticle in the final state 	has energy well above $T$. In Ref.~\cite{karzig_energy_2010} the definition of the energy relaxation rate equivalent to Eqs.~(\ref{eq:tau}) and (\ref{eq:taudef}) was applied to account for the effect of finite temperature on the relaxation due to the processes of type (a). This means that out of the two right-moving quasiparticles with energies much greater than $T$, the one with the higher momentum was identified as a new state of the initial quasiparticle.

The scattering rate entering Eq.~(\ref{eq:tau}) can be found using Fermi's golden rule, where the matrix element is obtained in the second-order perturbation theory in the interaction given by Eq.~(\ref{eq:V}) \cite{lunde_three-particle_2007,matveev_relaxation_2020}. In order to take advantage of the conservation laws, we express the momenta $p_1$, $p_2$, and $p_3$ in terms of the new variables $P$, $\mathcal E$, and $\alpha$ as	
\begin{align}\label{eq:p1p2p3}
p_j=\frac{1}{3}P-2\sqrt{\frac{m\mathcal{E}}{3}}\cos\biggl(\alpha-\frac{2\pi j}{3}\biggr),\quad j=1,2,3.
\end{align} 
Here $P=p_1+p_2+p_3$ is the total momentum of three particles, while $\mathcal{E}=\varepsilon_{p_1}+\varepsilon_{p_2}+\varepsilon_{p_3}-P^2/6m$ is their total energy in the center-of-mass frame \footnote{Equivalently, $\mathcal{E}=[(p_1-p_2)^2+(p_1-p_3)^2+(p_2-p_3)^2]/6m$.}. There are analogous formulas for the primed momenta. The conservation laws dictate that collisions do not affect $P$ and $\mathcal E$, thus only changing the angle variable, $\alpha\to \alpha'$. This observation dictates the general form of the three-particle scattering matrix element
\begin{align}\label{eq:W}
W_{p_1,p_2,p_3}^{p_1',p_2',p_3'}=\Theta(\mathcal{E},\alpha,\alpha')\delta(\mathcal{E}-\mathcal{E}')\delta_{P,P'}.
\end{align}
Starting with a general expression for $W_{p_1,p_2,p_3}^{p_1',p_2',p_3'}$ \cite{lunde_three-particle_2007,matveev_relaxation_2020}, after a somewhat tedious calculation we obtain Eq.~(\ref{eq:W}) with 
\begin{gather}\label{eq:Theta}
\Theta=\frac{2592\pi e^8}{\hbar L^4 \mathcal E^2} \ln^2\biggl(\frac{\hbar^2}{m w^2 \mathcal E}\biggr) \frac{f(\alpha+\alpha')+f(\alpha-\alpha')}{[\cos(3\alpha)-\cos(3\alpha')]^2},\\
f(\theta)=\Bigg[\sum_{j=1}^{3} \sin\biggl(\frac{\theta}{2}+\frac{2\pi j}{3}\biggr) \ln \left|\sin\biggl(\frac{\theta}{2}+\frac{2\pi j}{3}\biggr)\right|\Bigg]^2.
\end{gather}
This result applies to any three-particle scattering process, provided that $\ln({\hbar^2}/m w^2 \mathcal E)\gg 1$. The latter condition takes the forms $\hbar/wp_F \gg 1$ and $\hbar v_F/w\epsilon \gg 1$ for the processes of types (a) and (b), respectively. Here $p_F=\sqrt{2m\mu}$ is the Fermi momentum.

We begin our evaluation of the relaxation rate of a quasiparticle with the energy $\epsilon=\varepsilon_p-\mu$ via type (b) processes by analyzing Eq.~(\ref{eq:tau}). The distribution functions at low temperature severely constrain the configurations of momenta which give significant contribution to $\dot{\epsilon}$. In the zero temperature limit we have $p_2, p_3 \to p_F$  corresponding to $\mathcal{E}^*=\epsilon^2/6\mu$ and $\alpha^*=5\pi/3$, see Eq.~(\ref{eq:p1p2p3}). We account for the deviations of $p_2$, $p_3$, $p_2'$, and $p_3'$ from $p_F$ and of $p_1'$ from $p_1$ at finite temperature in the leading order in small parameters $\varrho=(\mathcal{E}-\mathcal{E}^*)/\mathcal{E}^*$, $\sigma=\alpha-\alpha^*$, and $\sigma'=\alpha'-\alpha^*$.
The function (\ref{eq:Theta}) is only weakly dependent on $\varrho$, which we can therefore neglect, leading to
\begin{align}\label{eq:Thetatypei}
\Theta={}&\frac{4608\pi e^8\mu^2}{\hbar L^4\epsilon^4}\ln^2\left(\frac{\hbar v_F}{w\epsilon}\right)\!\left[\frac{\ln^2|\sigma-\sigma'|}{(\sigma+\sigma')^2} + \frac{\ln^2|\sigma+\sigma'|}{(\sigma-\sigma')^2}\right].
\end{align}
Equation (\ref{eq:Thetatypei}) is singular at $\sigma=\pm\sigma'$, which corresponds to the nullification of the energy denominators in the initial expression of the second-order perturbation theory for the scattering rate (\ref{eq:W}). For type (b) processes, these singularities lead to a divergent quasiparticle decay rate, defined by omitting the energy difference $(\varepsilon_{p_1'}-\varepsilon_p)$ in the right-hand side of Eq.~(\ref{eq:tau}). However, the energy relaxation rate given by Eqs.~(\ref{eq:taudef}) and (\ref{eq:tau}) is well defined.

We are now in a position to evaluate the rate of quasiparticle energy change $\dot\epsilon$ using Eq.~(\ref{eq:tau}). Converting the sum into an integral over the variables $P,\varrho,\sigma$, and their primed versions, we first perform the integrations that involve the $\delta$-functions and then integrate over $\varrho$. The remaining integral over $\sigma$ and $\sigma'$ is an antisymmetric function and thus nullifies the rate if one approximates $n_{p_1'}$ by $n_p$. Accounting for the leading-order deviation in the distribution function of $p_1'$  results in a term proportional to $\sigma^2-\sigma'^2$ \footnote{See Supplemental Material for the details.}. In combination with the energy difference in Eq. (\ref{eq:tau}), also proportional to $\sigma^2-\sigma'^2$, it regularizes the singularities arising from Eq.~(\ref{eq:Thetatypei}). For the resulting relaxation rate we eventually obtain \cite{Note4}
\begin{align}\label{eq:ratefinal}
\frac{1}{\tau_b}=\frac{64\pi}{5\hbar} \left(\frac{e^2}{\hbar v_F}\right)^4 \ln^2 \left(\frac{\hbar v_F}{w \epsilon}\right) \ln^2\left(\frac{\epsilon}{T}\right) \frac{\mu^3 T^4}{\epsilon^6}.
\end{align}
Equation (\ref{eq:ratefinal}) is our main result. We now compare it with the energy relaxation rate due to the competing type (a) processes  \cite{karzig_energy_2010}.

Unlike the processes shown in Fig.~\ref{fig:hole}(b), the ones of Fig.~\ref{fig:hole}(a) contribute to quasiparticle relaxation even at $T=0$. In this case the quasiparticle decay rate is well defined despite the singularities in Eq.~(\ref{eq:Thetatypei}). It is given by \cite{karzig_energy_2010}
\begin{align}\label{eq:rateT=0}
\frac{1}{\tau_a}\sim \frac{1}{\hbar} \left(\frac{e^2}{\hbar v_F}\right)^4 \ln^2 \left(\frac{\hbar}{w p_F}\right) \ln^2\left(\frac{\mu}{\epsilon}\right) \frac{\epsilon^2}{\mu}.
\end{align}
The evaluation of the decay rate at finite temperatures is plagued by the singularities of Eq.~(\ref{eq:Thetatypei}). Instead, the energy relaxation rate (\ref{eq:taudef}) can be studied. At $T\gg \epsilon^2/\mu$ the result
\begin{align}\label{eq:particle2-1Tbigmain}
\frac{1}{\tau_a}\sim\frac{1}{\hbar}\left(\frac{e^2}{\hbar v_F}\right)^4 \ln^2\left(\frac{\hbar}{w p_F}\right) \ln^2\left(\frac{\mu}{\epsilon}\right) T
\end{align}
was found in Ref.~\cite{karzig_energy_2010}. It is worth mentioning that at $T=0$ the energy relaxation rate has the same form as the quasiparticle decay rate (\ref{eq:rateT=0}), albeit with a different numerical prefactor \cite{Note4}. A comparison of Eqs.~(\ref{eq:ratefinal})\,--\,(\ref{eq:particle2-1Tbigmain}) shows that the quasiparticles with energies $\epsilon\gg \sqrt{T\mu}$ decay with the rate (\ref{eq:rateT=0}), while at $T\ll \epsilon\ll\sqrt{T\mu}$ our result (\ref{eq:ratefinal}) gives the dominant contribution \cite{Note1}. For unscreened Coulomb interaction we conclude that the contribution (\ref{eq:particle2-1Tbigmain}) is always subdominant.

We now briefly discuss the relaxation of a hole, which represents the absence of a fermion in the Fermi sea. Because they propagate at speeds below
the Fermi velocity, holes are stable excitations at zero temperature. At nonzero temperatures they drift toward the Fermi level as a result of scattering off other excitations. At $\epsilon_h \gg T$, where $\epsilon_h=\mu-\varepsilon_p$ denotes the energy of the hole, the corresponding rate of energy change and the relaxation rate can be obtained from the expressions analogous to Eqs.~(\ref{eq:tau}) and (\ref{eq:taudef}). In Eq.~(\ref{eq:tau}) one should properly order the summation indices and replace the quasiparticle distribution function $n_p$, the dispersion $\varepsilon_p$, and $\epsilon$, respectively, by the corresponding quantities for holes, $1-n_p$, $-\varepsilon_p$, and $\epsilon_h$. For type (b) processes, the evaluation parallels the one for particles and results in the relaxation rate (\ref{eq:ratefinal}), with $\epsilon$ replaced by $\epsilon_h$.

\begin{figure}
	\includegraphics[width=0.5\columnwidth]{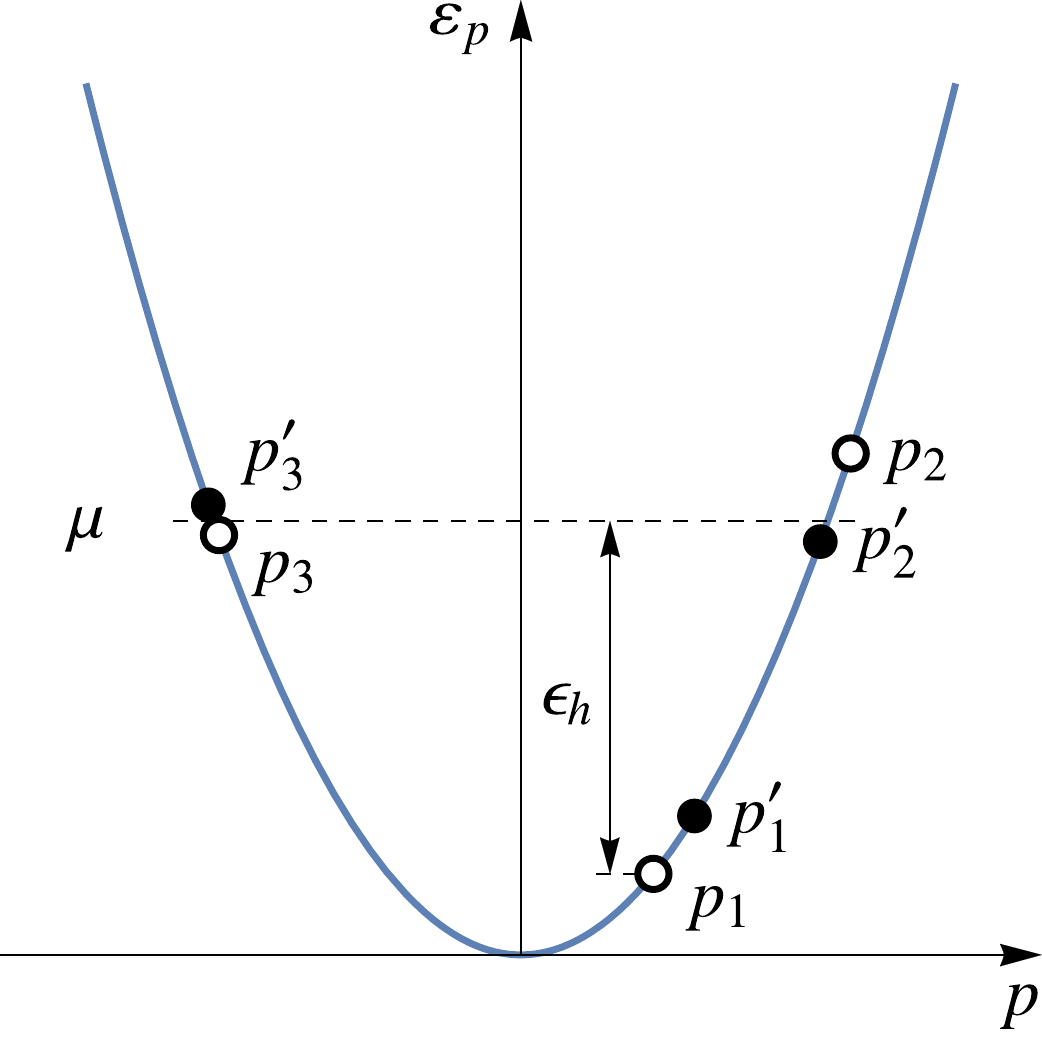}
	\caption{The dominant scattering mechanism that contributes to the relaxation of a deep hole, i.e., at $\epsilon_h\gg\sqrt{T \mu}$.} \label{fig2}
\end{figure}

Holes can also relax due to processes that involve quasiparticles near both Fermi points, see Fig.~\ref{fig2}. Since the left-moving pair has a characteristic momentum $|p_3-p_3'|\lesssim T/v_F$, from Eq.~(\ref{eq:exact}) we find the energy change of the hole
\begin{align}\label{eq:holech}
\Delta\epsilon_h= \frac{p_1'+p}{2m}(p_1'-p_1)\lesssim \textrm{min}\left(\epsilon_h,\frac{T\mu}{\epsilon_h}\right).
\end{align}
At $\epsilon_h\gg\sqrt{T \mu}$, we have $\Delta\epsilon_h\ll\epsilon_h$, i.e., the hole loses a small fraction of its energy in a three-particle collision. For such deep holes we can define the rate of energy change $\dot\epsilon_h$ and the relaxation rate $\tau_h^{-1}$ using the approach analogous to that of Eqs.~(\ref{eq:tau}) and (\ref{eq:taudef}) for particle-like excitations. The rate of energy change of a hole is given by \cite{Note4}
\begin{align}\label{eq:holedeepedot}
\dot{\epsilon}_h={}&-\frac{1}{\hbar} \left(\frac{e^2}{\hbar v_F}\right)^4    
\ln^2\left(\frac{\hbar}{w p_F}\right)T^2 F\left(\frac{p}{p_F}\right),
\end{align}
where
\begin{align}\label{eq:F}
F(a)={}&\frac{2}{\pi} \left(\ln\frac{1-a^2}{4}+a\ln\frac{1+a}{1-a}\right)^2\frac{a^2}{(1-a^2)^3}.
\end{align}
Equation (\ref{eq:holedeepedot}) is valid for deep holes, i.e., for $\epsilon_h=\mu-\varepsilon_p\gg \sqrt{T \mu}$. In the special case $\sqrt{T\mu} \ll \epsilon_h \ll \mu$ corresponding to deep holes near the Fermi level, from Eq.~(\ref{eq:holedeepedot}) we find $\tau_h^{-1}\propto\mu T^2/\epsilon_h^2$ \cite{Note1}. This result is consistent with the corresponding expression given in Ref.~\cite{karzig_energy_2010}. We note that Eq.~(\ref{eq:holedeepedot}) was obtained to leading order in low temperature, which limits its applicability to $p\gg\sqrt{mT}$.  An accurate expression for smaller $p$ is obtained by multiplying Eq.~(\ref{eq:holedeepedot}) by $1+mT/p^2$ \cite{Note4}.

Equation (\ref{eq:particle2-1Tbigmain}) for the energy relaxation rate due to the processes shown in Fig.~\ref{fig:hole}(a) \cite{karzig_energy_2010} and our Eq.~(\ref{eq:ratefinal}) for relaxation due to the processes of Fig.~\ref{fig:hole}(b) are applicable to both particles and holes. In particular, they apply to shallow holes with energies in the range
$T\ll\epsilon_h\ll\sqrt{T \mu}$ \cite{Note4}. Comparing the obtained results, we find that the relaxation of deep holes occurs primarily due to processes shown in Fig.~\ref{fig2}. In this case Eq.~(\ref{eq:holedeepedot}) gives the dominant contribution to their rate of energy change. In contrast, the relaxation of shallow holes with energies in the range $T\ll\epsilon_h\ll\sqrt{T \mu}$ is controlled by processes shown in Fig.~\ref{fig:hole}(b). Their relaxation rate is given by Eq.~(\ref{eq:ratefinal}) with $\epsilon$ replaced by $\epsilon_h$, while the corresponding rate of energy change follows from Eq.~(\ref{eq:taudef}).

In this paper we studied quasiparticles with energies $\epsilon\gg T$. This condition was important for the applicability of the approach based on Eq.~(\ref{eq:tau}), which assumes that the initial state of  momentum $p$ is not thermally populated. At $\epsilon\sim T$ one must account for the effect of thermal population of the state $p$, which can be achieved in a Boltzmann equation description. An order of magnitude estimate of the typical relaxation rate of the distribution function in the latter approach can be obtained by extrapolating the rate (\ref{eq:ratefinal}) to $\epsilon\sim T$,
\begin{align}\label{eq:lowT}
\frac{1}{\tau_b}\sim \frac{1}{\hbar} \left(\frac{e^2}{\hbar v_F}\right)^4 \ln^2 \left(\frac{\hbar v_F}{w \mspace{1mu}T}\right) \frac{\mu^3}{T^2}.
\end{align}
Unlike most other systems of fermions, in our case the relaxation rate increases at lower temperatures.  This can be attributed to the long-range nature of Coulomb interaction, which results in a singularity of the interaction potential at zero momentum and thus enhances scattering at small momentum transfer \cite{matveev_relaxation_2020}.

The fact that the relaxation rate (\ref{eq:lowT}) increases at $T\to0$
raises an important question of the applicability of the picture of
fermionic quasiparticles and holes used in this paper.  Indeed, at
sufficiently low temperature one may expect to reach the regime where
the standard assumption $\hbar/\tau_b\ll T$ is violated.  In this case
the uncertainty of the energy of a typical quasiparticle
$\delta\epsilon\sim\hbar/\tau_b$ is comparable to or larger than the
energy itself, $\epsilon\sim T$, and the quasiparticles are no longer
well defined.  In addition, in systems of weakly interacting
spin-$\frac12$ fermions the well-known phenomenon of spin-charge
separation \cite{Giamarchi,dzyaloshinskii_correlation_1974} results in
breakdown of the fermionic quasiparticle description.  As a result,
only the excitations with sufficiently high energies can be treated as
quasiparticles \cite{karzig_energy_2010}.  For an excitation with
energy $\epsilon\sim T$ in a system with long-range interactions the
condition of Ref.~\cite{karzig_energy_2010} can be presented in the
form $T\gg p_F V(T/v_F)/\hbar$.  For Coulomb interactions this yields
\begin{align}\label{eq:condition}
T\gg T^*=\mu \frac{e^2}{\hbar v_F}\ln\left(\frac{\hbar^2 v_F}{w p_F e^2}\right).
\end{align}
Our results are obtained under the assumptions that the interactions
are weak, $e^2/\hbar v_F \ll 1$, and the width of the channel is
small, $w p_F/\hbar \ll 1$.  In this case Eq.~(\ref{eq:condition})
ensures that the condition $\hbar/\tau_b\ll T$ is also satisfied.

In summary, we have studied the rate of energy relaxation for quasiparticles and holes in a weakly-interacting one-dimensional system of fermions with Coulomb repulsion. Compared to the case of screened interaction, we have found that scattering processes shown in Fig.~\ref{fig:hole}(b) lead to a dramatic enhancement of the quasiparticle relaxation rate at low energies, $\tau^{-1}\propto \epsilon^{-6}$ at $T\ll\epsilon\ll\sqrt{T\mu}$, see Fig.~\ref{fig3}. A similar enhancement also holds for shallow holes. For deep holes we have obtained their energy relaxation at arbitrary momenta, see Eq.~(\ref{eq:holedeepedot}).

Work at Argonne National Laboratory was supported by the US Department of Energy, Office of Science, Basic Energy Sciences, Materials Sciences and Engineering Division.


%

\onecolumngrid
\newpage
\setcounter{equation}{0}
\setcounter{figure}{0}

\renewcommand{\theequation}{S\arabic{equation}}
\renewcommand{\thepage}{S\arabic{page}}
\renewcommand{\thesection}{S\arabic{section}}
\renewcommand{\thetable}{S\arabic{table}}
\renewcommand{\thefigure}{S\arabic{figure}}

\renewcommand{\bibnumfmt}[1]{[{\normalfont S#1}]}


\setcounter{page}{1}

\begin{center}
	{\large\textbf{\TITLE}
		\\\vskip 5pt
		\normalsize{--Supplemental Material--}
		\\\vskip 5pt
	}
	
	Zoran Ristivojevic$^1$ and K. A. Matveev$^2$	\vskip 0.5mm
	\textit{\small$^1$Laboratoire de Physique Th\'{e}orique, Universit\'{e} de Toulouse, CNRS, UPS, 31062 Toulouse, France}\\
	\textit{\small $^2$Materials Science Division, Argonne National Laboratory, Argonne, Illinois 60439, USA}

\end{center}
\vskip 1.5pt

\section{I.~~Evaluation of the relaxation rate (\ref{eq:ratefinal}) controlled by type (b) processes}

Here we provide the details of the evaluation of the relaxation rate (\ref{eq:taudef}) for the processes shown in Fig.~\ref{fig:hole}(b). Substituting  Eq.~(\ref{eq:tau}) expressed in the symmetric form in the definition (\ref{eq:taudef}) yields
\begin{align}\label{eq:tauoriginal}
\frac{1}{\tau}={}&\frac{1}{24}\sum_{{p_1,p_2,p_3\atop p_1',p_2',p_3'}} \frac{\varepsilon_p-\text{max}(\varepsilon_{p_1'},\varepsilon_{p_2'}, \varepsilon_{p_3'})}{\epsilon} W_{p_1,p_2,p_3}^{p_1',p_2',p_3'} n_{p_2}n_{p_3}(1-n_{p_1'})(1- n_{p_2'})(1- n_{p_3'})\delta_{p,p_1}.
\end{align}
We have transformed the restricted summation of Eq.~(\ref{eq:tau}) into the unrestricted one accounting for the combinatorial factor $1/12$ to compensate the summation over nondistinct quantum states. We then transform the sum (\ref{eq:tauoriginal}) into an integral and convert to the new variables introduced by Eq.~(\ref{eq:p1p2p3}) using
\begin{align}\label{jacobian}
dp_1 dp_2 dp_3=\frac{m}{\sqrt{3}} dP d\mathcal{E}d\alpha.
\end{align}
After performing the trivial integrations over the $\delta$-functions, we find
\begin{align}\label{eq:taueval}
\frac{1}{\tau}=\frac{L^4 m^2}{384\pi^4 \hbar^4}\frac{g_p}{n_p} \int_{0}^{\infty}d\mathcal{E}\int_{0}^{2\pi}d\alpha \int_{0}^{2\pi}d\alpha'\, \frac{\varepsilon_p-\text{max}(\varepsilon_{p_1'},\varepsilon_{p_2'}, \varepsilon_{p_3'})}{\epsilon} \Theta(\mathcal{E},\alpha,\alpha') g_{p_2} g_{p_3} g_{p_1'} g_{p_2'} g_{p_3'}.
\end{align}
Here $g_p=\sqrt{n_p(1-n_p)}=[2\cosh\left((\varepsilon_p-\mu)/2T\right)]^{-1}$, while the momenta are 
\begin{subequations}\label{eq:pjboth}
	\begin{align}\label{eq:pj}
	p_j={}&p+2\sqrt{\frac{m\mathcal{E}}{3}}\left[\cos(\alpha-2\pi/3)-\cos(\alpha-2\pi j/3) \right],\\
	\label{eq:pjp}
	p_j'={}&p+2\sqrt{\frac{m\mathcal{E}}{3}}\left[\cos(\alpha-2\pi/3)-\cos(\alpha'-2\pi j/3) \right],\quad j=1,2,3.
	\end{align}
\end{subequations}
In the limit of small temperature, both $|p_2-p_F|$ and $|p_3-p_F|$ should be small due to $g_{p_2}g_{p_3}$. At $T\to 0$ this occurs at $\alpha=5\pi/3$, corresponding to $p_2=p_3=p_F$. On the contrary, out of three terms $|p_j'-p_F|$, only two can vanish simultaneously. This occurs at three values $\alpha'=\pi/3,\pi,5\pi/3$, corresponding to different exchanges of the primed momenta. For the purpose of further evaluation of the rate we select one configuration, e.g.,  $\alpha'=5\pi/3$ and multiply the rate by 3. Accounting for small fluctuations in the arguments of $g$-functions (controlled by the temperature), we use the expressions 
\begin{gather}
p_{2,3}=p_F-\frac{\epsilon}{2v_F}\varrho\pm \frac{\epsilon}{\sqrt{3}v_F}\sigma,\quad
p_{2,3}'=p_F-\frac{\epsilon}{2v_F}\varrho\pm \frac{\epsilon}{\sqrt{3}v_F}\sigma',\quad
p_1'=p+\frac{\epsilon}{3v_F}(\sigma^2-\sigma'^2).\label{eq:expanded}
\end{gather}
Here $v_F=p_F/m$ is the Fermi velocity, and we introduced new variables $\sigma$, $\sigma'$, and $\varrho$ via
\begin{align}\label{wwwwww}
\alpha=\alpha^*+\sigma,\quad \alpha'=\alpha^*+\sigma',\quad \mathcal{E}=\mathcal{E}^*(1+\varrho).
\end{align}
For the scattering processes of type (b), at $T\ll\epsilon_h$ and the above choice of $\alpha'$ we should use $\varepsilon_p-\text{max}(\varepsilon_{p_1'},\varepsilon_{p_2'}, \varepsilon_{p_3'})\simeq \varepsilon_p-\varepsilon_{p_1'}$. Linearization of the spectrum $\varepsilon_p-\varepsilon_{p_1'}=v_F(p-p_1')$ together with the leading-order term $g_{p_{1'}}=g_p$ substituted into Eq.~(\ref{eq:taueval}) produces zero after integration, since the integrand is antisymmetric to the exchange of $\sigma$ and $\sigma'$. We thus need the subleading term, which is proportional to $\sigma^2-\sigma'^2$. It can arise either from nonlinearity of the spectrum, leading to a  multiplicative term on the order of $T^2/\mu\epsilon$ or the Taylor expansion $g_{p_1'}=g_p+g_p'(p_1'-p)+\ldots$, where the analogous term is on the order of $T/\epsilon$. We keep the latter contribution since it is  parametrically larger. Using the matrix element given in Eq.~(\ref{eq:Thetatypei}) we find
\begin{align}\label{eq:tausupp}
\frac{1}{\tau_b}=-\frac{8}{9\pi^3}\left(\frac{e^2}{\hbar v_F}\right)^4 \frac{\mu^3}{\hbar v_F\epsilon} \frac{g_p g_p'}{n_p}\ln^2\left(\frac{\hbar v_F} {w\epsilon}\right)\iiint_{-\infty}^{+\infty} d\varrho d\sigma d\sigma' (\sigma^2-\sigma'^2)^2 \left[\frac{\ln^2|\sigma-\sigma'|}{(\sigma+\sigma')^2}+\frac{\ln^2|\sigma+\sigma'|}{(\sigma-\sigma')^2}\right] g_{p_2} g_{p_3} g_{p_2'} g_{p_3'},
\end{align} 
where we should eventually substitute Eq.~(\ref{eq:expanded}). The integral in Eq.~(\ref{eq:tausupp}) can be evaluated analytically with logarithmic accuracy, resulting in $(48\pi^4/5) (T/\epsilon)^5 \ln^2(\epsilon/T)$. Using ${g_p g_p'}/{n_p}=-v_F/2T$ at $p\to p_F^{+}$ in the low-temperature regime $T\ll \varepsilon_p-\mu$, from Eq.~(\ref{eq:tausupp}) we then obtain Eq.~(\ref{eq:ratefinal}). 

It is worth mentioning that the scaling in Eq.~(\ref{eq:ratefinal}) can be schematically understood from Eq.~(\ref{eq:taueval}) as
\begin{align}
\frac{1}{\tau_b}\propto  \underbrace{\mathcal{E}^*\left(\frac{T}{\epsilon}\right)^3}_{\int d\mathcal{E} d\alpha d\alpha	'} \times \underbrace{\frac{1}{T^2\epsilon^2}}_{\textrm{from }\Theta} \times \underbrace{\frac{T^2}{\epsilon^2}}_{\frac{\varepsilon_{p_1}-\varepsilon_{p_1'}}{\epsilon}} \times \underbrace{\frac{T}{\epsilon}}_{\textrm{expansion of } g_{p_1'}}\propto\frac{T^4}{\epsilon^6}.
\end{align}
We note that our evaluation of $\tau_b^{-1}$  also applies for the relaxation of holes due to type (b) processes after the transformation where we replace in Eq.~(\ref{eq:taueval}) the quasiparticle distribution function $n_p$, the dispersion $\varepsilon_p$, and $\epsilon$, respectively, by the corresponding quantities for holes, $1-n_p$, $-\varepsilon_p$, and $\epsilon_h$.

\section{II.~~Evaluation of the relaxation rates (\ref{eq:rateT=0}) and (\ref{eq:particle2-1Tbigmain}) controlled by type (a) processes}

In this section we evaluate the relaxation rate as well as the decay rate of a quasiparticle with energy $\epsilon$  due to the processes shown in Fig.~\ref{fig:hole}(a) at $T=0$. We demonstrate that up to numerical coefficients both rates are given by Eq.~(\ref{eq:rateT=0}). We also reproduce the result for the relaxation rate at finite temperature (\ref{eq:particle2-1Tbigmain}), first obtained in Ref.~\cite{karzig_energy_2010}.

We consider a three-particle process involving a quasiparticle with momentum $p=p_1$ near $p_F$ and two additional fermions at $p_2$ and $p_3$ near $p_F$ and $-p_F$, respectively, see Fig.~\ref{fig:hole}(a). The deviations of $p_2$ and $p_3$ from $\pm p_F$ are controlled by the small parameter  $\epsilon/\mu \ll 1$. Here $\epsilon=\varepsilon_p-\mu\simeq v_F(p-p_F)\gg T$. We begin with Eq.~(\ref{eq:taueval}) rewritten as
\begin{align}
\frac{1}{\tau}={}&\frac{L^4 m^2 }{384\pi^4\hbar^4}  \int_{0}^{+\infty}d\mathcal{E} \int_0^{2\pi}d\alpha \int_0^{2\pi} d\alpha'\, \frac{\varepsilon_p-\text{max}(\varepsilon_{p_1'},\varepsilon_{p_2'}, \varepsilon_{p_3'})}{\epsilon}  \Theta(\mathcal{E},\alpha,\alpha') n_{p_2} n_{p_3} (1-n_{p_1'})(1-n_{p_2'})(1-n_{p_3'}),
\label{eq:rate2-1}
\end{align}
where the momenta are given by Eqs.~(\ref{eq:pjboth}). Unlike the case of type (b) processes, where all three particles in the final state are near the same Fermi point, this is not the case in Eq.~(\ref{eq:rate2-1}), leading to a minor ambiguity in the meaning of $\text{max}(\varepsilon_{p_1'},\varepsilon_{p_2'}, \varepsilon_{p_3'})$. Normally one would require all the momenta in the arguments of $\textrm{max}$ to be on the same branch as $p$. The error introduced here is negligible as the Fermi sea prevents the left-moving particle from being more than $T$ above the Fermi level, which is smaller than the typical energies of the right-moving particles in the final state. Since $p_2$ and $p_3$ are near $\pm p_F$, we can consider the configuration $p_3<0<p_2$ multiplying the rate (\ref{eq:rate2-1}) by 2. The configuration $p_2=-p_3=p_F$ corresponds to
\begin{align}\label{eq:E2-1T}
\mathcal{E}^*=\frac{p^2+3p_F^2}{3m}\simeq \frac{8\mu}{3},\quad \cos(\alpha^*)=\frac{p+3p_F}{2\sqrt{3m\mathcal{E}^*}},\quad \sin(\alpha^*)=\frac{p_F-p}{2\sqrt{m\mathcal{E}^*}}.
\end{align} 
Each of the three primed momenta in the integrand of Eq.~(\ref{eq:rate2-1}) can be near $-p_F$, but only one is allowed by the conservation laws. We select $p_3'$ near $-p_F$ and multiply the rate (\ref{eq:rate2-1}) by 3. Accounting for the deviations around $\alpha^*\simeq -\sqrt{3}\epsilon/8\mu$ via Eq.~(\ref{wwwwww}), to leading order in small  $\varrho$, $\sigma$, and $\sigma'$, we find
\begin{subequations}
\begin{align}
p_1={}&p,\quad
&&p_1'=p+\frac{2p_F}{\sqrt{3}} (\sigma-\sigma')+\ldots,\\
p_2={}&p_F + \frac{4p_F}{\sqrt{3}}\sigma+\ldots,
&&p_2'=p_F + \frac{2p_F}{\sqrt{3}}(\sigma +\sigma')+\ldots,\\
p_3={}&-p_F - p_F\varrho+ \frac{3p_F-p}{\sqrt{3}}\sigma +p_F\sigma^2 +\ldots,\quad
&&p_3'=p_3-\frac{p_F-p}{\sqrt{3}}(\sigma-\sigma')-\frac{2p_F}{3}(\sigma^2-\sigma'^2)+\ldots.
\end{align}
\end{subequations}

We notice that unlike $p_3$ and $p_3'$, to leading order in $\epsilon/\mu \ll 1$ the momenta $p_1'$, $p_2$, and $p_2'$ do not depend on $\varrho$.  This enables us to rewrite Eq.~(\ref{eq:rate2-1}) as
\begin{align}
\frac{1}{\tau_a}\simeq{}&\frac{  L^4 m^2 \mathcal{E}^*}{64\pi^4\hbar^4}\iint_{-\pi/3}^{+\pi/3} d\sigma d\sigma'\, \frac{\varepsilon_p-\text{max}(\varepsilon_{p_1'},\varepsilon_{p_2'}, \varepsilon_{p_3'})}{\epsilon}  \Theta(\mathcal{E}^*,\alpha^*+\sigma,\alpha^*+\sigma') n_{p_2}  (1-n_{p_1'})(1-n_{p_2'})\int_{-1}^{+\infty}d\varrho\, n_{p_3} (1-n_{p_3'}).
\label{eq:rate2-1lowT}
\end{align}
Notice that in the latter integral we can extend the integration over the whole real axis, resulting in 
\begin{align}
\int_{-\infty}^{+\infty}d\varrho\, n_{p_3} (1-n_{p_3'}) \simeq\frac{4}{3\mathcal{E}^*}\times\begin{cases}
\Delta \theta(\Delta),\quad &T\ll \Delta,\\
T,\quad &\Delta\ll T,
\end{cases}
\quad\text{where}\quad \Delta=v_F(p_3-p_3').
\end{align}
Since the characteristic energy change of the right-moving pairs is  on the order of $\epsilon$, while it is parametrically smaller, $\Delta\sim\epsilon^2/\mu \ll \epsilon$, for the left-moving pair, we can neglect the temperature effects on the right movers as long as $T\ll\epsilon$. We thus use
\begin{align}
n_{p_2}  (1-n_{p_1'})(1-n_{p_2'})=\theta(-\sigma)\theta\left(\sigma-\sigma'-2\alpha^*\right) \theta(\sigma+\sigma'). 
\end{align}
The Heaviside functions impose for the integration boundaries
$\alpha^*<\sigma<0$ and $\quad -\sigma< \sigma'<-2\alpha^*+\sigma$ in Eq.~(\ref{eq:rate2-1lowT}). Since in the present case the deviations $\sigma$ and $\sigma'$ can be on the order of $\alpha^*$, it is convenient to change variables and use $\alpha$ and $\alpha'$ in Eq.~(\ref{eq:rate2-1lowT}). In this case the integration boundaries are
\begin{align}
2\alpha^*<\alpha<\alpha^*,\quad -(\alpha-2\alpha^*)<\alpha'<\alpha-2\alpha^*.
\end{align}
Since $\alpha\sim \alpha'\ll 1$, we should consider Eq.~(\ref{eq:Theta}) in this regime. It is given by
\begin{align}
\Theta(\mathcal E^*,\alpha,\alpha')={}&\frac{2^5 \pi}{\hbar L^4}\frac{e^8}{(\mathcal E^*)^2} \ln^2\left(\frac{\hbar^2}{mw^2\mathcal E^*}\right) \left[\frac{\ln^2|\alpha-\alpha'|}{(\alpha+\alpha')^2}+ \frac{\ln^2|\alpha+\alpha'|}{(\alpha-\alpha')^2}\right],\quad \mathcal{E}^*=\frac{8\mu}{3}.
\label{eq:Theta2-1-zeroTapp}
\end{align}
The remaining terms in Eq.~(\ref{eq:rate2-1lowT}) are
\begin{gather}
\Delta\simeq\frac{4\mu}{3}(\alpha'-\alpha)(-\alpha'-\alpha)>0,\\
\frac{\varepsilon_p-\text{max}(\varepsilon_{p_1'},\varepsilon_{p_2'}, \varepsilon_{p_3'})}{\epsilon}\simeq\frac{p-\textrm{max}(p_1',p_2')}{p-p_F}\simeq \frac{4\mu}{\sqrt{3}\epsilon}\left[\alpha'-\alpha-2\alpha' \theta(\alpha')\right].
\end{gather}
In the latter expression we have used $\theta(p_1'-p_2')=\theta(-\alpha')$.
Using
\begin{align}\label{eqln}
\int_{2\alpha^*}^{\alpha^*}d\alpha \int_{-(\alpha-2\alpha^*)}^{\alpha-2\alpha^*} d\alpha' [\alpha'-\alpha-2\alpha'\theta(\alpha')](\alpha'-\alpha) (-\alpha-\alpha') \left[\frac{\ln^2(\alpha'-\alpha)}{(\alpha+\alpha')^2}+ \frac{\ln^2(-\alpha'-\alpha)}{(\alpha'-\alpha)^2}\right]
\simeq\frac{53-12\ln 2}{256\sqrt{3}}\frac{\epsilon^3}{\mu^3}\ln^2\left(\frac{\mu}{\epsilon}\right),
\end{align}
at $T\ll \epsilon^2/\mu$ we find the relaxation rate 
\begin{align}\label{lnln}
\frac{1}{\tau_a}=\frac{53-12\ln 2}{96\pi^3\hbar} \left(\frac{e^2}{\hbar v_F}\right)^4 \ln^2 \left(\frac{\hbar}{w p_F}\right) \ln^2\left(\frac{\mu}{\epsilon}\right) \frac{\epsilon^2}{\mu},
\end{align}
corresponding to Eq.~(\ref{eq:rateT=0}). As mentioned in the main text, at $T=0$ the quasiparticle decay rate is well defined. Omitting the factor $[\varepsilon_p-\text{max}(\varepsilon_{p_1'},\varepsilon_{p_2'}, \varepsilon_{p_3'})]/\epsilon$ in Eq.~(\ref{eq:rate2-1}) and using
\begin{align}
\int_{2\alpha^*}^{\alpha^*}d\alpha \int_{-(\alpha-2\alpha^*)}^{\alpha-2\alpha^*} d\alpha' (\alpha'-\alpha) (-\alpha-\alpha') \left[\frac{\ln^2(\alpha'-\alpha)}{(\alpha+\alpha')^2}+ \frac{\ln^2(-\alpha'-\alpha)}{(\alpha'-\alpha)^2}\right]
\simeq{}&\frac{9}{64}\frac{\epsilon^2}{\mu^2}\ln^2\left(\frac{\mu}{\epsilon} \right),
\end{align}
we obtain the quasiparticle decay rate in the form of Eq.~(\ref{lnln}) with the numerical prefactor $9/32\pi^3$. This is in agreement with Eq.~(1) of Ref.~\cite{karzig_energy_2010}.

At $\epsilon^2/\mu \ll T\ll\epsilon$, one can still take the zero-temperature limit in the distribution functions of the right movers in the initial expression (\ref{eq:rate2-1lowT}). Using
\begin{align}
\int_{2\alpha^*}^{\alpha^*}d\alpha \int_{-(\alpha-2\alpha^*)}^{\alpha-2\alpha^*} d\alpha' [\alpha'-\alpha-2\alpha'\theta(\alpha')] \left[\frac{\ln^2(\alpha'-\alpha)}{(\alpha+\alpha')^2}+ \frac{\ln^2(-\alpha'-\alpha)}{(\alpha'-\alpha)^2}\right]\simeq
\frac{\sqrt{3}(4\ln2 -1)}{8}\frac{\epsilon}{\mu}\ln^2\left(\frac{\mu}{\epsilon}\right),
\end{align}
we find the relaxation rate
\begin{align}\label{eq:particle2-1Tbig}
\frac{1}{\tau_a}=\frac{3(4\ln 2-1)}{4\pi^3}\frac{1}{\hbar}\left(\frac{e^2}{\hbar v_F}\right)^4 \ln^2\left(\frac{\hbar}{w p_F}\right) \ln^2\left(\frac{\mu}{\epsilon}\right) T,
\end{align}
	in agreement with Eq.~(2) of Ref.~\cite{karzig_energy_2010}. Omitting the numerical coefficient, we obtain Eq.~(\ref{eq:particle2-1Tbigmain}).
 
The results (\ref{lnln}) and (\ref{eq:particle2-1Tbig}) can be interpreted as follows. The three-body scattering matrix element is obtained from the two-particle interaction in second-order perturbation theory. For spin-$\frac{1}{2}$ fermions, the result contains an energy denominator proportional to $\epsilon$. In the decay rate, the corresponding scattering probability, which is proportional to $1/\epsilon^2$, enters multiplied by the available phase space volume for the three-particle scattering, $\nu$. For type~(a) processes, the typical energy of the counter-propagating particle-hole pair is $\epsilon^2/\mu$, while it is $\epsilon$ for the other two pairs. This yields  $\nu\propto \epsilon^4/\mu$ and the decay rate $\tau^{-1}\propto \epsilon^2/\mu$. For temperatures in the range $\epsilon^2/\mu \ll T \ll \epsilon$ the phase space contribution of the counter-propagating particle-hole pair is replaced by $T$, leading to $\nu\propto \epsilon^2 T$ and $\tau^{-1}\propto T$. To obtain the logarithmic energy dependence in the latter result, a careful calculation is required.

\section{III.~~Phase-space expectation values}

In this section we study the phase space expectation values of various quantities that describe the three particle scattering of hole excitations. We find two characteristic regimes that represent shallow and deep holes. In the former case the scattering matrix element is the same as for the particles, see Eq.~(\ref{eq:Theta2-1-zeroTapp}); for deep holes it is evaluated below, see Eq.~(\ref{eq:Thetaexp}).

We begin by considering the expectation value defined by
\begin{align}
\langle M(\ldots) \rangle=\frac{\sum \delta_{p,p_1} \delta(\mathcal{E}-\mathcal{E}')\delta_{P,P'} M(\ldots) (1-n_{p_2}) (1-n_{p_3}) n_{p_1'} n_{p_2'} n_{p_3'}}{\sum \delta_{p,p_1} \delta(\mathcal{E}-\mathcal{E}')\delta_{P,P'}(1-n_{p_2}) (1-n_{p_3}) n_{p_1'} n_{p_2'} n_{p_3'}},
\end{align}
where the summation is over $p_1$, $p_2$, $p_3$, $p_1'$, $p_2'$, and $p_3'$, while $M(\ldots)$ is an arbitrary function of the summation indices. Converting the sum into an integral and expressing the Fermi distribution functions in terms of $g_p=\sqrt{n_p(1-n_p)}$ (which requires $T>0$), we obtain
\begin{align}\label{eq:meanM}
\langle M(\ldots) \rangle=\frac{\int_{0}^{\infty}d\mathcal{E}\iint_{0}^{2\pi}d\alpha d\alpha' M(\ldots) g_{p_2}g_{p_3}g_{p_1'}g_{p_2'}g_{p_3'}}{ \int_{0}^{\infty}d\mathcal{E}\iint_{0}^{2\pi}d\alpha d\alpha'  g_{p_2}g_{p_3}g_{p_1'}g_{p_2'}g_{p_3'} },
\end{align}
where the momenta are given by Eq.~(\ref{eq:pjboth}). We now concentrate on the configuration of momenta describing the processes shown in Fig.~\ref{fig2} with a hole of momentum $p$ in the initial state and the two holes on the opposite sides that are in the vicinity of $\pm p_F$.
Out of two configurations that at $T=0$ are $p_2=\pm p_F$ and $p_3=\mp p_F$,
we select, e.g., the case $p_3<0<p_2$ multiplying the integrands in the numerator and the denominator of Eq.~(\ref{eq:meanM}) by 2. While the latter multiplicative factor is irrelevant for Eq.~(\ref{eq:meanM}), it will become important for Eq.~(\ref{eq:rateholef}). For $p_3=-p_F$ and $p_2=p_F$ we have
\begin{align}\label{eq:alpha2-1}
\mathcal{E}^*=\frac{p^2+3p_F^2}{3m},\quad \cos(\alpha^*)=\frac{p+3p_F}{2\sqrt{p^2+3p_F^2}},\quad \sin(\alpha^*)=\frac{\sqrt{3}(p_F-p)}{2\sqrt{p^2+3p_F^2}}.
\end{align} 
Equation (\ref{eq:alpha2-1}) is very similar to Eq.~(\ref{eq:E2-1T}), but in the present case $p$ covers a wider range $-p_F<p<p_F$. At $T>0$ we should account for small deviations of $\alpha$, $\alpha'$, and $\mathcal E$ from their values (\ref{eq:alpha2-1}) via Eq.~(\ref{wwwwww}). Due to the specific choice of $\sigma'$ out of three equivalent ones, we multiply the integrands in the numerator and the denominator of Eq.~(\ref{eq:meanM}) by $3$. Again, this is irrelevant here, but needed for the evaluation of Eq.~(\ref{eq:rateholef}). We then have
\begin{subequations}\label{eq:p1p2p3hole}
	\begin{align}
	p_1={}&p, &&p_1'=p+\frac{2}{\sqrt{3}}p_F (\sigma-\sigma'),\\
	p_2={}&p_F + \frac{p_F-p}{2}\varrho+ \frac{3p_F+p}{\sqrt{3}}\sigma, &&p_2'=p_2-\frac{p_F+p}{\sqrt{3}}(\sigma-\sigma'),\\
	p_3={}&-p_F - \frac{p_F+p}{2}\varrho+ \frac{3p_F-p}{\sqrt{3}}\sigma, &&p_3'=p_3-\frac{p_F-p}{\sqrt{3}}(\sigma-\sigma').\label{eq:p3hole}
	\end{align}
\end{subequations}
We now simplify $g$-functions by linearizing the spectrum around the Fermi points in Eq.~(\ref{eq:meanM}). We note that for $p$ near $\pm p_F$ we can use
\begin{align}\label{eq:gpexp}
g_p\simeq\frac{1}{2} \sech\left(\frac{v_F(p\mp p_F)}{2T}\right).
\end{align}
The approximation done in Eq.~(\ref{eq:gpexp}) is justified provided $v_F|p\pm p_F|\ll \sqrt{T \mu}$, since in that case the subleading terms are small. We use the expression (\ref{eq:gpexp}) for all $g$-functions except $g_{p_1'}$. This yields
\begin{align}\label{eq:average}
\langle M(\ldots) \rangle=\frac{\iiint_{-\infty}^{+\infty} dx dy dz  M(\ldots)\sech(x) \sech\boldsymbol(x-\left(1+p/p_F\right)z\boldsymbol)\sech(y) \sech\boldsymbol(y-\left(1-{p}/{p_F}\right)z\boldsymbol)g_{p+{4Tz}/{v_F} }}{\iiint_{-\infty}^{+\infty} dx dy dz  \sech(x) \sech\boldsymbol(x-\left(1+p/p_F\right)z\boldsymbol)\sech(y) \sech\boldsymbol(y-\left(1-{p}/{p_F}\right)z\boldsymbol)g_{p+{4Tz}/{v_F} }},
\end{align}
where instead of the original variables $\varrho,\sigma,\sigma'$ we have introduced
\begin{align}\label{eq:xyz}
x=\frac{v_F}{2T}(p_2-p_F),\quad y=\frac{v_F}{2T}(p_3+p_F),\quad z=\frac{v_F}{4T}(p_1'-p).
\end{align}
The corresponding transformation is
\begin{subequations}\label{jacc}
	\begin{gather}
\sigma=\frac{T}{\sqrt{3} \mathcal{E}^*}\left[\left(1+\frac{p}{p_F}\right)x +\left(1-\frac{p}{p_F}\right)y\right],\quad
\sigma'=\sigma-\frac{\sqrt{3}T}{\mu }z,\quad
\varrho=\frac{2T}{3\mathcal{E}^*}\left[\left(3-\frac{p}{p_F}\right)x-\left( 3+\frac{p}{p_F}\right)y\right],\\
d\varrho d\sigma d\sigma'=\frac{2T^3}{\mu^2  \mathcal{E}^*} dx dy dz.
\end{gather}
\end{subequations}
In order to find average values $\langle p_2\rangle$, \ldots, $\langle p_3'\rangle$ it is thus sufficient to evaluate $\langle x\rangle$, $\langle y\rangle$, and $\langle z\rangle$. We use Eq.~(\ref{eq:average}) and
\begin{align}
&\int_{-\infty}^{+\infty} dx \sech(x)\sech(x-a z)=\frac{2a z}{\sinh(a z)},\quad
&\int_{-\infty}^{+\infty} dx x \sech(x)\sech(x-a z)=\frac{a^2 z^2}{\sinh(a z)},
\end{align}
to find
\begin{gather}
\langle x\rangle=\frac{1}{2}\left(1+\frac{p}{p_F}\right)\frac{L_3}{L_2},\quad
\langle y\rangle=\frac{1}{2}\left(1-\frac{p}{p_F}\right)\frac{L_3}{L_2},\quad
\langle z\rangle=\frac{L_3}{L_2},\\
\langle\sigma\rangle=\frac{T}{\sqrt{3} \mathcal{E}^*}\left(1+\frac{p^2}{p_F^2}\right)\frac{L_3}{L_2},\quad \langle\sigma'\rangle=\frac{T}{\sqrt{3} \mathcal{E}^*}\left(1+\frac{p^2}{p_F^2}-\frac{3\mathcal{E}^*}{\mu} \right)\frac{L_3}{L_2},\quad \langle\varrho\rangle=\frac{4T}{3\mathcal{E}^*}\frac{p}{p_F}\frac{L_3}{L_2}.
\end{gather}
Here we have introduced
\begin{subequations}\label{eq:Lj}
\begin{align}\label{eq:Ljint}
L_j(p)=\frac{1}{g_p}\int_{-\infty}^{+\infty} dz \frac{z^j g_{p+4Tz/v_F} }{\sinh\boldsymbol((1+p/p_F)z\boldsymbol)\sinh\boldsymbol((1-p/p_F)z\boldsymbol)},
\end{align}
where
\begin{align}\label{eq:gg}
\frac{g_{p+4Tz/v_F}}{g_p}={}&\frac{1}{\cosh\left(\frac{2p}{p_F}z+\frac{2T}{\mu} z^2\right)-\tanh\left(\frac{\epsilon_h}{2T}\right) \sinh\left(\frac{2p}{p_F}z+\frac{2T}{\mu} z^2\right)}. 
\end{align}
\end{subequations}
The integral (\ref{eq:Ljint}) is non-singular because for the holes the momentum is assumed to be in the range $-p_F<p<p_F$. We can evaluate analytically $L_j(p)$  in two regimes of interest, (i) shallow hole, $T\ll\epsilon_h\ll\sqrt{T\mu}$, and (ii) deep hole, $\sqrt{T\mu}\ll \epsilon_h<\mu$.

In the regime (i), the linearization $\sinh\boldsymbol((1-p/p_F)z\boldsymbol)\simeq \epsilon_h z/2\mu$ is justified since the integrand is appreciable only at $|z|\lesssim \epsilon_h/T$; otherwise it decays very rapidly. The same condition enabled us to neglect the terms proportional to $z^2$ in Eq.~(\ref{eq:gg}).
We then find
\begin{align}\label{eq:Ljlowp}
L_j(p)\simeq {}&\frac{p_F}{p_F- p} \int_{-\infty}^{+\infty} dz \frac{z^{j-1}}{\sinh(2z)}\frac{1}{\cosh(2z)- \tanh\left(\frac{\epsilon_h}{2T}\right)\sinh(2z)}.
\end{align}
The latter integral can be evaluated exactly. In the cases of interest we obtain
\begin{align}
L_2(p)\simeq {}&\frac{1}{8}\frac{\mu}{\epsilon_h} \left[\pi^2+\left(\frac{\epsilon_h}{T}\right)^2\right],\\
L_3(p)\simeq{}& L_2(p) \frac{\epsilon_h}{6T},
\end{align}
which will be useful for our study of the hole near the Fermi points.

In the regime (ii) $\sqrt{T\mu}\ll \epsilon_h<\mu$ we can approximate $\tanh(\epsilon_h/T)\simeq 1$, leading to
\begin{align}\label{eq:approx}
\frac{g_{p+4Tz/v_F}}{g_p}\simeq e^{2pz/p_F}e^{2Tz^2/\mu}.
\end{align}
It is useful to introduce
\begin{subequations}
\label{eq:curlyLj}
\begin{align}
\mathcal{L}_{j}(a)={}&2\int_{0}^{+\infty} dz \frac{z^{j} \sinh(2a z)} {\sinh\boldsymbol((1+a)z\boldsymbol)\sinh\boldsymbol((1-a)z\boldsymbol)}=\frac{j!\zeta(j+1)}{2^{j-1}}\left[\frac{1}{(1-a)^{j+1}}-\frac{1}{(1+a)^{j+1}}\right],\\
\mathcal{\bar L}_{j}(a)={}&2\int_{0}^{+\infty} dz \frac{z^{j} \cosh(2a z)} {\sinh\boldsymbol((1+a)z\boldsymbol)\sinh\boldsymbol((1-a)z\boldsymbol)}=\mathcal{L}_{j}(a) +8\int_0^\infty dz \frac{z^j}{\left[e^{2(1+a)z}-1\right]\left[1-e^{-2(1-a)z}\right]}.
\end{align}
\end{subequations}
The former integral is evaluated using $\sinh(2 a z)=\sinh\boldsymbol((1+a)z-(1-a)z\boldsymbol)$, which enabled us to simplify the integrand and then perform the integration. We note that the integrand in Eq.~(\ref{eq:curlyLj}) is considerable at $z\lesssim 1/(1\mp a)\lesssim \mu/\epsilon_h$. Even at such large $z$ the exponent of the second exponential function in Eq.~(\ref{eq:approx}) is at most on the order of $T\mu/\epsilon_h^2\ll 1$. It can be therefore neglected in the leading order result. In the regime (ii) we have thus obtained 
$L_{j}(p)\simeq\mathcal{L}_{j}(p/p_F)$ for odd $j$ and $L_{j}(p)\simeq\mathcal{\bar L}_{j}(p/p_F)$ for even $j$. For $\epsilon_h \ll \mu$ we have $\mathcal{\bar L}_{j}(a)\simeq \mathcal{L}_{j}(a)$. A comparison of the expressions for $L_j(p)$ in the regime (i) given by Eq.~(\ref{eq:Ljlowp}) and for the regime (ii) given by Eq.~(\ref{eq:curlyLj}) enables us to find that at $\epsilon_h\sim \sqrt{T\mu}$ they are of the same order, as expected.

For $p\to p_F^-$ and $\epsilon_h\gg\sqrt{T\mu}$, we have $\mathcal{E}^*=8\mu/3$ and $L_3(p)/L_2(p)\sim \mu/\epsilon_h$. This leads to $\langle x\rangle\simeq\langle z\rangle\sim \mu/\epsilon_h$, $\langle y\rangle\sim 1$,
and therefore
\begin{gather}
\langle p_1'-p\rangle\sim\frac{T\mu }{v_F\epsilon_h},\quad
\langle p_2-p_F\rangle=\frac{\langle p_2-p_2'\rangle}{2}\sim\frac{T\mu}{v_F\epsilon_h},\quad
\langle p_F+p_3\rangle=\frac{\langle p_3-p_3'\rangle}{2}\sim\frac{T}{v_F},\\
\langle \sigma \rangle\sim
\langle \sigma' \rangle\sim
\langle\varrho\rangle\sim\frac{T}{\epsilon_h}.
\end{gather}

In the case of $p\to p_F^-$ and $\epsilon_h\ll\sqrt{T\mu}$, we have obtained $L_3(p)/L_2(p)=\epsilon_h/6T$. This leads to $\langle x\rangle\simeq\langle z\rangle\sim\epsilon_h/T$, $\langle y\rangle\sim \epsilon_h^2/T\mu$,
and therefore
\begin{gather}
\langle p_1'-p\rangle\sim\frac{\epsilon_h}{v_F},\quad
\langle p_2-p_F\rangle=\frac{\langle p_2-p_2'\rangle}{2}\sim\frac{\epsilon_h}{v_F},\quad
\langle p_F+p_3\rangle=\frac{\langle p_3-p_3'\rangle}{2}\sim\frac{\epsilon_h^2}{v_F \mu },\\
\langle \sigma \rangle\sim
\langle \sigma' \rangle\sim
\langle\varrho\rangle\sim\frac{\epsilon_h}{\mu}.
\end{gather}
We can now see that the linearization of the $g$-functions [cf.~Eq.~(\ref{eq:gpexp})] for the momenta $p_2$, $p_2'$, $p_3$, and $p_3'$ was indeed justified. 

The average energy change of the hole in a single scattering event is
\begin{align}\label{eq:deltaeps}
\Delta \epsilon_h=\langle \varepsilon_{p_1'}-\varepsilon_p\rangle\simeq \frac{p}{m}\langle p_1'-p\rangle=4T\frac{p}{p_F}\frac{{L}_3(p)}{{ L}_2(p)}=\begin{cases}
\frac{8\pi^2}{5}\frac{p^2}{p_F^2}T, & p\ll p_F,\\
\frac{2\pi^4}{15\zeta(3)} \frac{\mu}{\epsilon_h}T, &\sqrt{T\mu} \ll \epsilon_h \ll \mu,\\
\frac{2\epsilon_h}{3}, &T \ll \epsilon_h\ll \sqrt{T\mu}.
\end{cases}
\end{align}
The latter expression is consistent with the estimate made in Eq.~(\ref{eq:holech}). Equation (\ref{eq:deltaeps}) shows that $\Delta \epsilon_h$ increases as its momentum $p$ increases from small values toward $p_F$. It also justifies the definition introduced after Eq.~(\ref{eq:Lj}) of a shallow hole, with  $T\ll\epsilon_h\ll\sqrt{T \mu}$, and a deep hole, for which $\epsilon_h \gg \sqrt{T\mu}$.  For shallow holes $\Delta\epsilon_h \sim \epsilon_h$, whereas for deep ones $\Delta\epsilon_h \ll \epsilon_h$.

For $p\to p_F^-$,  from Eq.~(\ref{eq:alpha2-1}) we obtain $\alpha^*=\sqrt{3}\epsilon_h/8\mu$.  Since $\langle\sigma\rangle\sim \langle\sigma'\rangle\sim \epsilon_h/\mu$, the deviations of $\alpha$ and $\alpha'$ from $\alpha^*$ are on the same order as $\alpha^*$ for shallow holes, $\epsilon_h\ll \sqrt{T\mu}$. 
This has an important consequence for the evaluation of $\Theta(\mathcal{E}^*,\alpha,\alpha')$ [see Eq.~(\ref{eq:Theta})] that controls the relaxation rate. In the case a shallow hole one should use the expression given by Eq.~(\ref{eq:Theta2-1-zeroTapp}). However, in the case of a deep hole, $\epsilon_h\gg\sqrt{T\mu}$, one should expand Eq.~(\ref{eq:Theta}) in the small deviations $\sigma\sim\sigma'\ll \alpha^*$. This yields
\begin{gather}\label{eq:Thetaexp}
\Theta(\mathcal{E}^*,\alpha=\alpha^*+\sigma,\alpha'=\alpha^*+\sigma')=\frac{18\pi}{\hbar L^4} e^8\ln^2\left(\frac{\hbar^2}{mw^2\mathcal{E}^*}\right) \frac{\left(\ln\frac{p_F^2-p^2}{4p_F^2}+\frac{p}{p_F}\ln\frac{p_F+p}{p_F-p}\right)^2}{(\mu-\varepsilon_p)^2 (\sigma-\sigma')^2},\quad \mathcal{E}^*=\frac{p^2+3p_F^2}{3m}.
\end{gather}
The result (\ref{eq:Thetaexp}) applies at arbitrary $p$ such that $\epsilon_h\gg \sqrt{T\mu}$. It is obtained from Eq.~(\ref{eq:Theta}) using \begin{subequations}\label{eq:cosexp}
\begin{gather}
\cos(3\alpha^*+3\sigma)-\cos(3\alpha^*+3\sigma')=-\frac{3p_F(p_F^2-p^2)}{(m\mathcal{E}^*)^{3/2}}(\sigma-\sigma')+O(\sigma^2-\sigma'^2),\\
f(\alpha+\alpha')+f(\alpha-\alpha')=f(2\alpha^*)+O(\sigma+\sigma'),\quad 
f(2\alpha^*)=\frac{p_F^2}{4m\mathcal{E}^*}\left(\ln\frac{p_F^2-p^2}{4p_F^2}+\frac{p}{p_F}\ln\frac{p_F+p}{p_F-p}\right)^2.
\end{gather}
\end{subequations}
The neglected terms in Eqs.~(\ref{eq:cosexp}) are subleading at $\epsilon_h\gg\sqrt{T\mu}$.

\section{IV.~~Evaluation of the energy relaxation of holes}

In this section we study the energy relaxation of holes. For deep holes we obtain the rate of energy change (\ref{eq:holedeepedot}) at arbitrary momenta. For shallow holes we show that the rate corresponding to the process shown in Fig.~\ref{fig:hole}(a) is given by Eq.~(\ref{eq:particle2-1Tbig}) for quasiparticle relaxation, with $\epsilon$ replaced with $\epsilon_h$.

The relaxation rate of a hole of momentum $p$ in the one-dimensional Fermi gas can be obtained from an expression similar to Eq.~(\ref{eq:tau}) where one should replace $n_p$, $\varepsilon_p$, and $\epsilon$, respectively, by the corresponding quantities for holes, $1-n_p$, $-\varepsilon_p$, and $\epsilon_h$. Analogously to Eq.~(\ref{eq:taueval}) we then find
\begin{align}
\dot{\epsilon}_h={}&-\frac{L^4 m^2 }{384\pi^4\hbar^4} \frac{g_p}{1-n_p}\int_{0}^{+\infty}d\mathcal{E} \int_0^{2\pi}d\alpha \int_0^{2\pi} d\alpha'\,\left[\text{min}(\varepsilon_{p_1'},\varepsilon_{p_2'}, \varepsilon_{p_3'})-\varepsilon_p\right]\Theta(\mathcal{E},\alpha,\alpha')  g_{p_2} g_{p_3} g_{p_1'}g_{p_2'}g_{p_3'},
\label{eq:rateholef}
\end{align}
where the momenta are given by Eqs.~(\ref{eq:pjboth}). As usual, we assume that $\epsilon_h = \mu - \varepsilon_p \gg T$.

The evaluation of Eq.~(\ref{eq:rateholef}) for scattering processes with quasiparticles near both Fermi points closely follows the one of Eq.~(\ref{eq:meanM}). For the states near the Fermi points, we linearize the dispersion that enters into the occupation numbers, which enter through $g_p=\sqrt{n_p(1-n_p)}$, see Eq.~(\ref{eq:gpexp}). We must keep the full form of $g_{p_1'}$ in the case of a deep hole. This leads to
\begin{align}\label{eq87}
\dot{\epsilon}_h={}&-\frac{m^2L^4}{1024\pi^4\hbar^4} \frac{g_p^2}{1-n_p}\mathcal{E}^* \int_{-1}^{+\infty} d\varrho \iint_{-\pi/3}^{+\pi/3}d\sigma d\sigma' \left[\text{min}(\varepsilon_{p_1'},\varepsilon_{p_2'}, \varepsilon_{p_3'})-\varepsilon_p\right]  \Theta\boldsymbol{(}\mathcal{E}^*(1+\varrho),\alpha^*+\sigma,\alpha^*+\sigma'\boldsymbol{)} \notag\\
&\times \sech\left(\frac{v_F(p_2-p_F)}{2T}\right) \sech\left(\frac{v_F(p_2'-p_F)}{2T}\right)\sech\left(\frac{v_F(p_3+p_F)}{2T}\right) \sech\left(\frac{v_F(p_3'+p_F)}{2T}\right)\frac{g_{p_1}}{g_p}.
\end{align}
We notice that $g_p^2/(1-n_p)=n_p\simeq 1$ at $T\ll \epsilon_h$. In the following we evaluate Eq.~(\ref{eq87}) for the cases of deep and shallow holes.

In the case of a deep hole, $\epsilon_h\gg\sqrt{T\mu}$, we use $\text{min}(\varepsilon_{p_1'},\varepsilon_{p_2'}, \varepsilon_{p_3'})-\varepsilon_p\simeq\varepsilon_{p_1'}-\varepsilon_{p}$ since $|\mu-\varepsilon_{p_2'}|, |\mu-\varepsilon_{p_3'}|\lesssim T$, while $|\mu-\varepsilon_{p_1'}|\sim \epsilon_h\gg T$. We then substitute Eq.~(\ref{eq:Thetaexp}) in the expression (\ref{eq87}) and use the variables (\ref{eq:xyz}), with the corresponding transformation (\ref{jacc}). It leads to
\begin{align}\label{eq:pro1}
\dot{\epsilon}_h={}&-\frac{3}{4\pi^3}\frac{1}{\hbar} \left(\frac{e^2}{\hbar v_F}\right)^4  \ln^2\left(\frac{\hbar^2}{mw^2\mathcal{E}^*}\right) \left(\ln\frac{p_F^2-p^2}{4p_F^2}+\frac{p}{p_F}\ln\frac{p_F+p}{p_F-p}\right)^2 \frac{T^2\mu}{\epsilon_h}\left[\frac{p}{p_F}{L}_1(p)+\frac{T}{\mu}L_2(p)\right],
\end{align}
where ${L}_j(p)$ is defined by Eq.~(\ref{eq:Lj}). The second term in the expression in brackets of Eq.~(\ref{eq:pro1}) is smaller by a factor $T/\epsilon_h$ for $p$ near $p_F$, as follows from Eq.~(\ref{eq:curlyLj}). However, for small $p$ the bracket becomes $\pi^2(2\varepsilon_{p}+T)/3\mu$. At $p\gg \sqrt{mT}$, Eq.~(\ref{eq:pro1}) leads to Eq.~(\ref{eq:holedeepedot}), where we use $\ln^2(\hbar^2/mw^2\mathcal{E}^*)\simeq 4\ln^2(\hbar/w p_F)$. At $p \to p_F^-$, the expression in parentheses in Eq.~(\ref{eq:pro1}) becomes asymptotically equal $(1-p/p_F)\ln(1-p/p_F)$, while ${L}_1(p)\simeq \pi^2/6(1-p/p_F)^2$. The corresponding relaxation rate (\ref{eq:taudef}) of the deep hole then becomes
\begin{align}\label{eq:th}
\frac{1}{\tau_{h}}={}&\frac{1}{2\pi\hbar} \left(\frac{e^2}{\hbar v_F}\right)^4  \ln^2\left(\frac{\hbar}{w p_F}\right) \ln^2\left(\frac{\mu}{\epsilon_h}\right)\frac{T^2\mu}{\epsilon_h^2}.
\end{align}

We now consider the energy relaxation of a shallow hole, $T\ll\epsilon_h\ll \sqrt{T\mu}$, with the  momentum $p$ near $p_F$. In the variables introduced by Eq.~(\ref{eq:xyz}) we obtain
\begin{align}
\text{min}(\varepsilon_{p_1'},\varepsilon_{p_2'}, \varepsilon_{p_3'})-\varepsilon_p\simeq 4T\, \textrm{min}\left(z,\frac{\epsilon_h}{4T}+\frac{x}{2}-z\right).
\end{align}
Substituting Eq.~(\ref{eq:Theta2-1-zeroTapp}) in the expression (\ref{eq87}) we then obtain within the logarithmic accuracy
\begin{subequations}
\begin{align}\label{eq:pro}
\dot{\epsilon}_h=- \frac{3 J}{16\pi^3\hbar} \left(\frac{e^2}{\hbar v_F}\right)^4  \ln^2\left(\frac{\hbar}{wp_F}\right) \ln^2\left(\frac{\mu}{\epsilon_h}\right) {T\epsilon_h},
\end{align}
where we have introduced
\begin{align}
J=\frac{T}{\epsilon_h} \iiint_{-\infty}^{+\infty} dx dy dz\, \textrm{min}\left(z,\frac{\epsilon_h}{4T}+\frac{x}{2}-z\right) \left[\frac{1}{z^2}+\frac{1}{\left(\frac{\epsilon_h}{4T}+\frac{x}{2}-z\right)^2}\right]\sech (x)\sech(x-2z)\sech^2(y) \frac{g_{p+4Tz/v_F}}{g_p}.
\end{align} 
\end{subequations}
The integration over $y$ is straightforward. In the remaining integral we linearize the argument in $g$-functions, which is allowed for shallow holes, as discussed near Eq.~(\ref{eq:gpexp}). After introducing the new variables $u=4T x/\epsilon_h$ and $w=4T z/\epsilon_h$, we find
\begin{align}
J=\frac{1}{2}\int du dw\, \textrm{min}\left(w,1+\frac{u}{2}-w\right) \left[\frac{1}{w^2}+\frac{1}{\left(1+\frac{u}{2}-w\right)^2}\right] \sech \left(\frac{\epsilon_h}{4T}u\right) \sech \left(\frac{\epsilon_h}{4T}(u-2w)\right)\frac{\cosh\left(\frac{\epsilon_h}{2T} \right)}{\cosh\left(\frac{\epsilon_h}{2T}(1-w)\right)}.
\end{align}
In the regime $T\ll \epsilon_h$ the product of the hyperbolic functions simplifies into $4$ at $0<u<2w$, $0<w<1$, and $0$ otherwise. The integral then becomes elementary with the result $J=4(4\ln2-1)$. The relaxation rate $\tau_{h}^{-1}=-\dot{\epsilon}_h/\epsilon_h$ for shallow holes due to type (a) processes that follows from Eq.~(\ref{eq:pro}) then becomes identical to our previous expression (\ref{eq:particle2-1Tbig}) (as well as the corresponding one of Ref.~\cite{karzig_energy_2010}) that was derived for particles, provided one replaces $\epsilon$ by $\epsilon_h$. However, it is a subdominant contribution since type (b) processes lead to a larger rate that is given by Eq.~(\ref{eq:ratefinal}), with $\epsilon$ replaced by $\epsilon_h$.

\end{document}